\documentclass[twocolumn]{aastex63}

\newcommand\kms{km~s$^{-1}$}

\received{\today}
\revised{\today}
\accepted{}

\submitjournal{ApJ}

\shorttitle{Time Variability Discovered in Tycho's SNR}
\shortauthors{Matsuda et al.}

\graphicspath{{./}{./figures/}}
\begin{document}

\title{Discovery of Year-Scale Time Variability from Thermal X-ray Emission in Tycho's Supernova Remnant}

\correspondingauthor{Masamune Matsuda}
\email{matsuda.masamune.38a@kyoto-u.jp}

\author[0000-0002-7393-2234]{Masamune Matsuda}
\affiliation{Department of Physics, Kyoto University,
Kitashirakawa Oiwake, Sakyo, Kyoto,
Kyoto 606-8502, Japan}

\author[0000-0003-1518-2188]{Hiroyuki Uchida}
\affiliation{Department of Physics, Kyoto University,
Kitashirakawa Oiwake, Sakyo, Kyoto,
Kyoto 606-8502, Japan}

\author[0000-0002-4383-0368]{Takaaki Tanaka}
\affiliation{Department of Physics, Konan University, 
8-9-1 Okamoto, Higashinada, Kobe,
Hyogo 658-8501, Japan}

\author[0000-0002-5092-6085]{Hiroya Yamaguchi}
\affiliation{Institute of Space and Astronautical Science (ISAS), Japan Aerospace Exploration Agency (JAXA),
3-1-1 Yoshinodai, Chuo, Sagamihara,
Kanagawa 252-5210, Japan}
\affiliation{Department of Physics, The University of Tokyo,
7-3-1 Hongo, Bunkyo,
Tokyo, 113-0033, Japan}

\author[0000-0002-5504-4903]{Takeshi Go Tsuru}
\affiliation{Department of Physics, Kyoto University,
Kitashirakawa Oiwake, Sakyo, Kyoto,
Kyoto 606-8502, Japan}

\begin{abstract}
Mechanisms of particle heating are crucial to understanding the shock physics in supernova remnants (SNRs).
However, there has been little information on time variabilities of thermalized particles so far. 
Here, we present a discovery of a gradually-brightening thermal X-ray emission found in Chandra data of Tycho's SNR obtained during 2000--2015.
The emission exhibits a knot-like feature (Knot1) with a diameter of $\simeq0.04$~pc located in the northwestern limb, where we also find localized H$\alpha$ filaments in an optical image taken with the Hubble Space Telescope in 2008.
The model with the solar abundance reproduces the spectra of Knot1, suggesting that Knot1 originates from interstellar medium; this is the first detection of thermal X-ray emission from swept-up gas found in Tycho's SNR.
Our spectral analysis indicates that the electron temperature of Knot1 has increased from $\sim0.30$~keV to $\sim0.69$~keV within the period between 2000 and 2015.
These results lead us to ascribe the time-variable emission to a small dense clump recently heated by the forward shock at the location of Knot1.
The electron-to-proton temperature ratio immediately downstream the shock ($\beta_{0}\equiv T_e/T_p$) is constrained to be $m_e/m_p\leq\beta_{0}\leq0.15$ to reproduce the data, indicating the collisionless electron heating with efficiency consistent with previous H$\alpha$ observations of Tycho and other SNRs with high shock velocities.
\end{abstract}

%%%%%%%%%%%%%%%%%%%%%%%%%%%%%%%%%%%%%%%%%%%%%%%%%%%%%%%%%%%%%%%%%%%%%%%%%%%%%%%%

\section{Introduction}\label{sec:intro}

Physics of collisionless shocks is an intriguing topic since they are involved with a number of unsettled problems, e.g., the evolution of magnetic turbulence, the electron heating mechanism, and the process of cosmic-ray acceleration.
One poorly understood process among them is collisionless electron heating although it is an important subject which might be related to the formation of the collisionless shocks.
While several pieces of observational evidence for the collisionless heating have been found in various astrophysical environments such as solar wind shocks \citep{schw88}, supernova remnants \citep[SNRs; e.g.,][]{lami96, ghav01, yama14}, and merging galaxy clusters \citep[e.g.,][]{mark05, russ12}, the detailed heating mechanism in collisionless shocks is still under debate.

%The temperature change behind the shock provides a clue to the elusive fundamental properties of the collisionless electron heating. 
The temperature change at the shock front
provides a clue to the elusive fundamental properties of the collisionless electron heating. 
%When the collisionless heating is negligible, the temperature downstream of the shock with velocity of $v_\mathrm{sh}$ is written as $kT_i=(3/16)m_i v_\mathrm{sh}^2$, where $k$ is Boltzmann constant, $m_i$ is the mass of particle species $i$. 
When electron heating occurs without collisionless process such as plasma wave heating via Buneman instabilities \citep[e.g.,][]{carg88} and lower hybrid wave heating \citep[e.g.,][]{lami00}, the temperature downstream of the shock with velocity of $v_\mathrm{sh}$ is written as $kT_i=(3/16)m_i v_\mathrm{sh}^2$, where $k$ is Boltzmann constant, $m_i$ is the mass of particle species $i$.
It follows that the particle temperature in the shock transition is proportional to its mass. 
Thus, the electron temperature is much smaller than the temperature of heavier ions. 
The electrons then receive thermal energy from the ions via Coulomb collisions, and the temperature gradually increases. 
On the other hand, when the collisionless heating is efficient, the electron temperature rises quickly in the shock transition and gradually rises via Coulomb collisions further downstream \cite[e.g.,][]{mack74,carg88}.
Direct measurement of these temperature changes can constrain the efficiency of collisionless heating in a shock transition.

Recent observations found year-scale time variabilities of synchrotron X-rays in small scales in shock waves of young SNRs: RX~J1713.7$-$3946 \citep{uchi07}, Cassiopeia~A \citep[Cas~A;][]{uchi08} and G330.2$+$1.0 \citep{bork18}.
Our previous studies also revealed similar year-scale spectral changes in one of the youngest and nearby Type Ia SNRs, Tycho's SNR \citep[hereafter, Tycho;][]{okun20,mats20}.
These studies provided us with important information on a real-time energy change of non-thermal particles.
On the other hand, time variabilities of thermal X-rays, which help us solve the problem of the heating mechanism of thermalized particles, have been less reported except for several examples on Cas~A \citep[e.g.,][]{patn07,patn14,ruth13} and SN~1987A \cite[e.g.,][]{sun21,ravi21}.

Tycho has bright synchrotron X-ray rims with thermalized ejecta, in which \citet{yama14} revealed evidence of collisionless heating based on Fe-K diagnostics.
Although thermal X-ray emission from Tycho has been detected only from the ejecta heated by the reverse shock \citep[e.g.,][]{hwan02}, some studies suggested that the forward shock interacted with dense materials as evidenced by H$\alpha$ observations \citep[e.g.][]{ghav00,lee10} and velocity measurement of X-ray shells \citep{tana21}.
These results imply a presence of ISM heated very recently by the forward shock.
We, therefore, search for the thermal X-ray radiation from forward-shocked ISM and then investigate its temperature evolution through observations of short-timescale thermal variability using multiple archival Chandra datasets.
Throughout this paper, we adopt $\simeq2.5$~kpc as the distance to Tycho \citep{zhou16}, and the statistical errors are quoted at the 1$\sigma$ level.
%%%%%%%%%%%%%%%%%%%%%%%%%%%%%%%%%%%%%%%%%%%%%%%%%%%%%%%%%%%%%%%%%%%%%%%%%%%%%%%%

\section{Observations \& Data Reduction}\label{sec:observation}

Tycho was observed with the Chandra X-ray Observatory using ACIS-S  in 2000 and ACIS-I  in 2003, 2007, 2009, and 2015. Table \ref{tab:tycho_obs} presents the observation log.
We reprocess all the data with the Chandra Calibration Database (CALDB) version 4.8.2.
For relative astrometry corrections, we align the coordinates of each observation to that of the dataset with ObsID\,$=$\,10095, which has the longest effective exposure time.
We first detect point sources in the field using the CIAO task \verb|wavdetect|. 
We then reprocess all the event files using the tasks \verb|wcs_match| and \verb|wcs_update|.
Because the accuracy of the frame alignment depends on the photon statistics, short time observations (ObsID: 8551, 10903, 10904, and 10906) are discarded for the above astrometry and used only for spectral analysis.

\begin{deluxetable}{rccc}
	\tablecaption{Observation Log \label{tab:tycho_obs}}
	\tablewidth{0pt}
	\tablehead{
	\colhead{ObsID} & \colhead{Start date} & \colhead{Effective exposure} & \colhead{Chip} \\
	\colhead{} & & \colhead{(ks)} & \colhead{}
	}
	\startdata
		115	  & 2000 Oct 01 & 49 & ACIS-S\\
		3837  & 2003 Apr 29 & 146 & ACIS-I\\
		7639  & 2007 Apr 23 & 109 & ACIS-I\\
		8551  & 2007 Apr 26 & 33 & ACIS-I\\
		10093 & 2009 Apr 13 & 118 & ACIS-I\\
		10094 & 2009 Apr 18 & 90 & ACIS-I\\
		10095 & 2009 Apr 23 & 173 & ACIS-I\\
		10096 & 2009 Apr 27 & 106 & ACIS-I\\
		10097 & 2009 Apr 11 & 107 & ACIS-I\\
		10902 & 2009 Apr 15 & 40 & ACIS-I\\
		10903 & 2009 Apr 17 & 24 & ACIS-I\\
		10904 & 2009 Apr 13 & 35 & ACIS-I\\
		10906 & 2009 May 03 & 41 & ACIS-I\\
		15998 & 2015 Apr 22 & 147 & ACIS-I\\
	\enddata
\end{deluxetable}
%%%%%%%%%%%%%%%%%%%%%%%%%%%%%%%%%%%%%%%%%%%%%%%%%%%%%%%%%%%%%%%%%%%%%%%%%%%%%%%%

\section{Analysis \& Results}\label{sec:ana}
\subsection{Imaging Analysis}\label{subsec:img}

\begin{figure*}
    \epsscale{0.9}
	\plotone{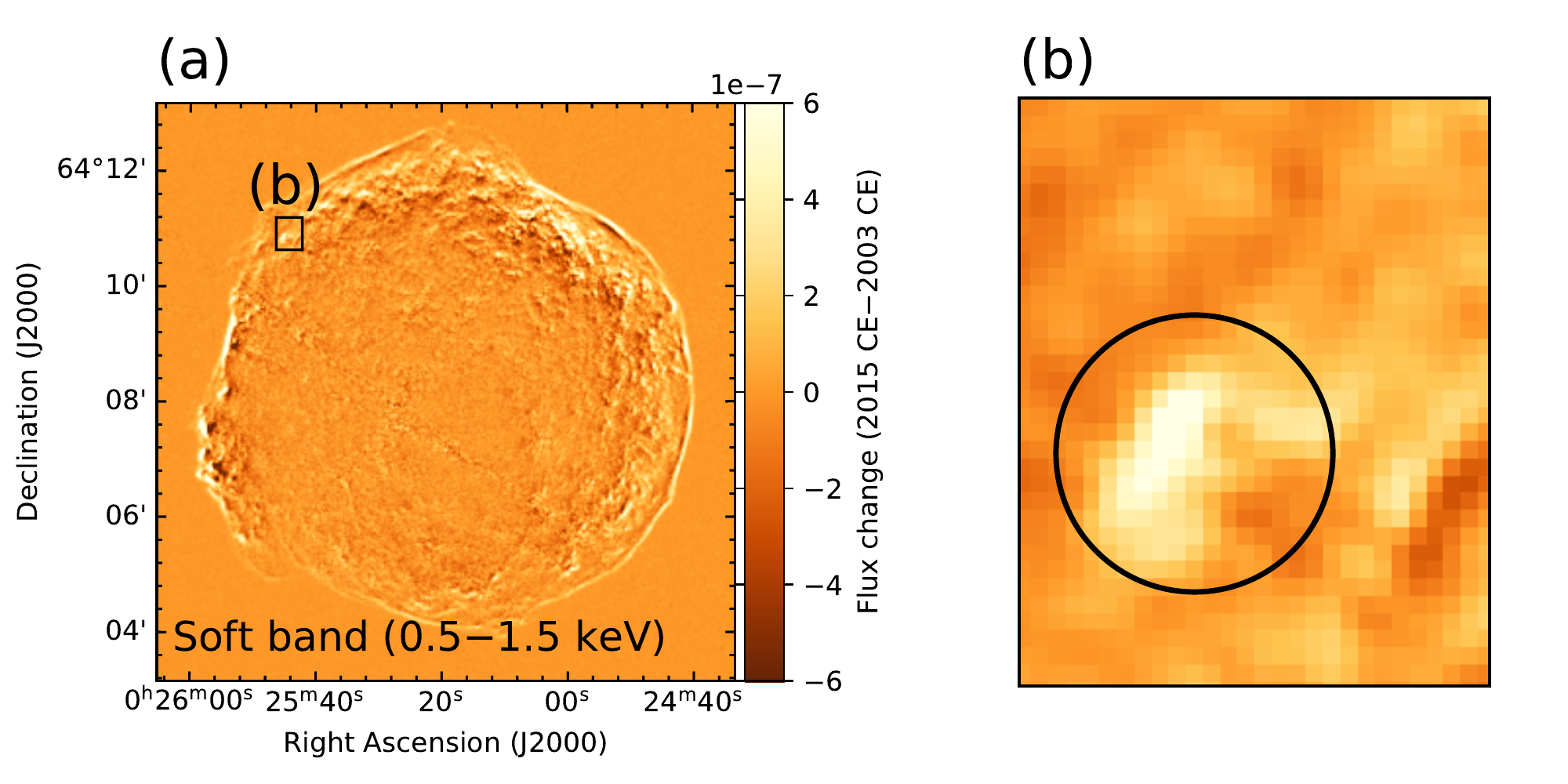}
	\caption{(a): Difference image of Tycho between 2003 and 2015 in the 0.5--1.5~keV band, where Ne-K$\alpha$ and Fe-L lines are dominated.
	The unit for the color scale is photons~s$^{-1}$~cm$^{-2}$. Knot1 is located in the box region.
	(b): Enlarged image of the box region indicated in panel~(a). The circle is the region where the flux changes significantly.
	\label{fig:tycho_diff}}
	\plotone{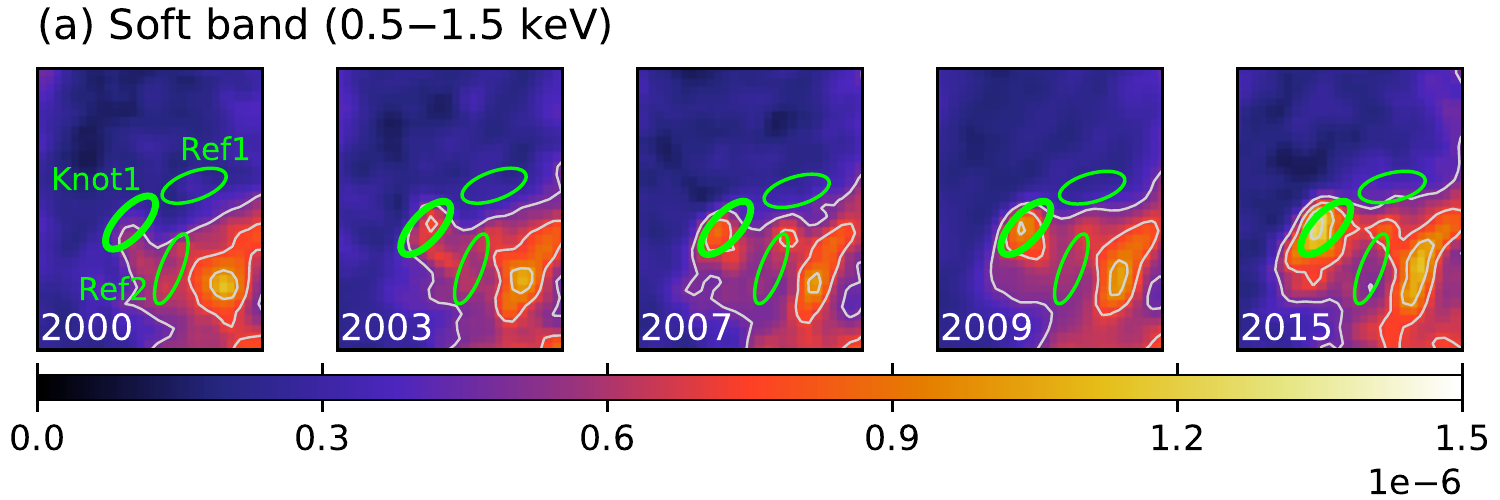}
	\plotone{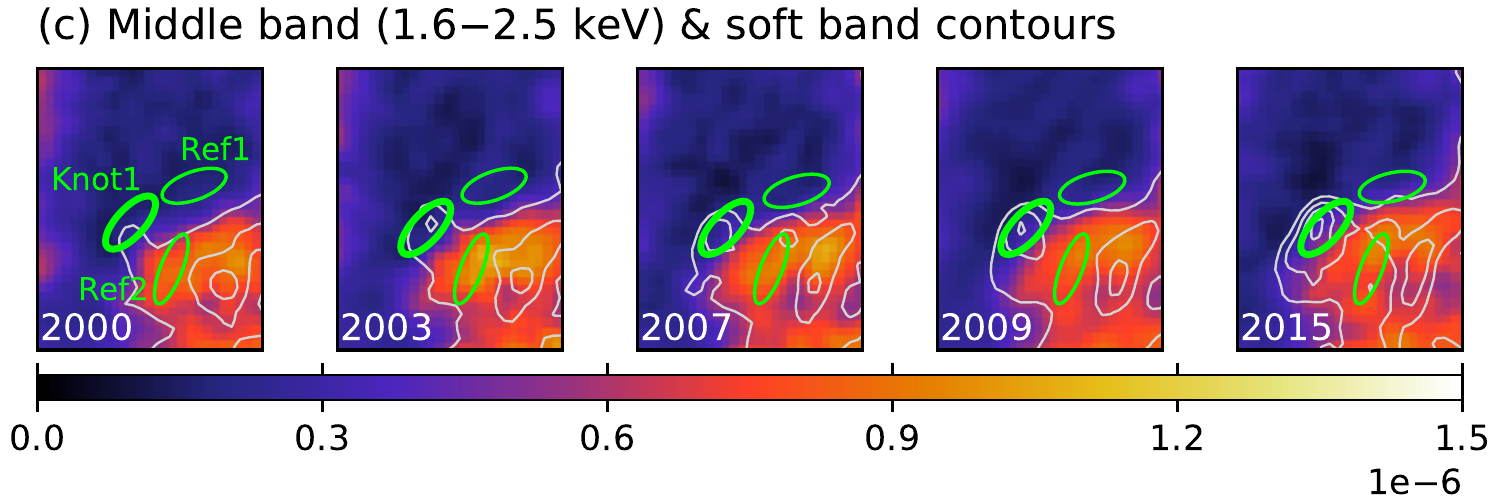}
	\plotone{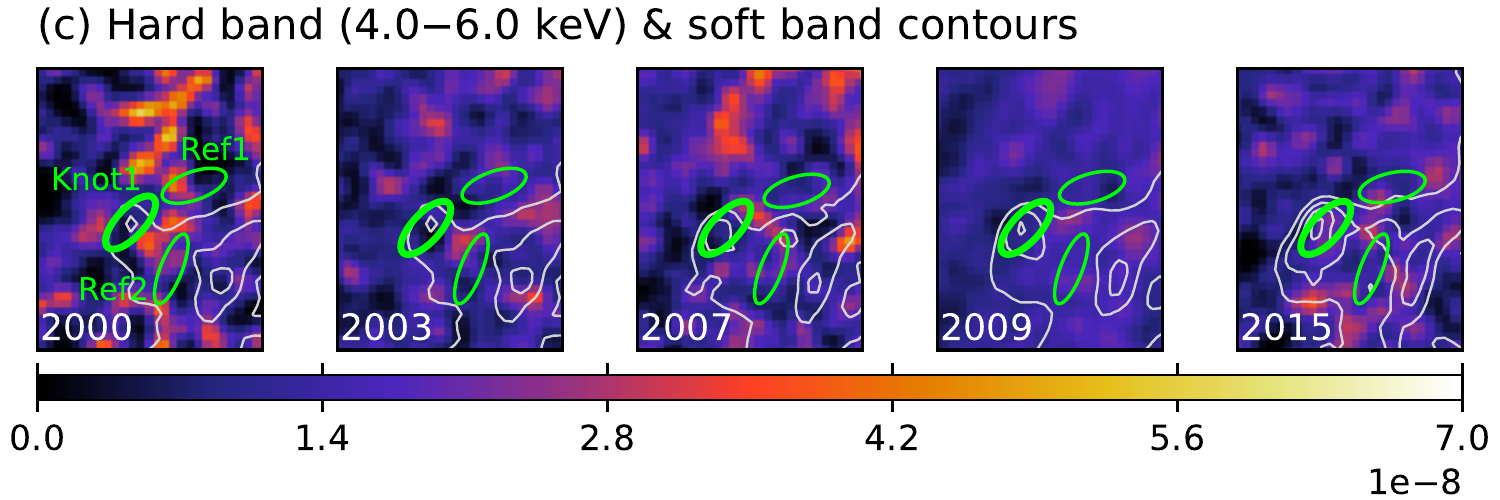}
	\caption{Soft (a), middle (b), and hard (c) band X-ray images around Knot1 taken in 2000, 2003, 2007, 2009, and 2015. 
	All the images are exposure-corrected.
	In all the panels, the unit for the color scale is photons~s$^{-1}$~cm$^{-2}$. 
	Contours represent the flux of the soft band X-rays.
	The green ellipses are the Knot1, Ref1, Ref2 regions used for spectral extraction, respectively.
	\label{fig:bandimg}}
\end{figure*}

To search for time variabilities of thermal emissions, we make a difference map by subtracting an exposure-corrected image taken in 2003 from one taken in 2015.
Since the thermal emission dominates in a soft band in most regions \citep[e.g.,][]{warr05, sato17}, we first focus on the lowest energy band (0.5--1.5~keV) as shown in Figure~\ref{fig:tycho_diff}.
The figure shows the flux change of thermal emission in the interior of the shell besides non-thermal emission at the shell.
Most features in the difference map show the flux increase and decrease, next to each other.
These features result from bright structures moving between 2000 and 2015 due to expanding ejecta and radial proper motions of the blast waves.
In the northeast, however, we discover a bright spot (hereafter, ``Knot1'') whose photon count monotonically increases over time with no signs of proper motion (panel~(b) of Figure~\ref{fig:tycho_diff}).

Figure \ref{fig:bandimg} shows  visual comparisons of flux images of Knot1 in the soft (0.5--1.5~keV), middle (1.6--2.5~keV), and hard (4.0--6.0~keV) bands.
For better statistics, the images in 2009 are created by adding together the observations of ObsIDs 10093, 10094, 10095, 10096, 10097, and 10902 after the astrometry corrections (see Section~\ref{sec:observation}).
We confirm in Figure~\ref{fig:bandimg}~(a) that Knot1 was gradually brightening from 2000 through 2015. 
On the other hand, Figure~\ref{fig:bandimg}~(b) shows no significant flux fluctuation but for the ejecta expansion, suggesting that Knot1 has a different origin from the middle band X-rays.
We also find that the synchrotron emission is relatively faint in Knot1 without any significant flux changes (Figure 2 c), unlike the ``stripe'' regions in the southwest \citep{okun20, mats20}.
%%%%%%%%%%%%%%%%%%%%%%%%%%%%%%%%%%%%%%%%%%%%%%%%%%%%%%%%%%%%%%%%%%%%%%%%%%%%%%%%

\begin{figure*}
    \epsscale{1.1}
	\plotone{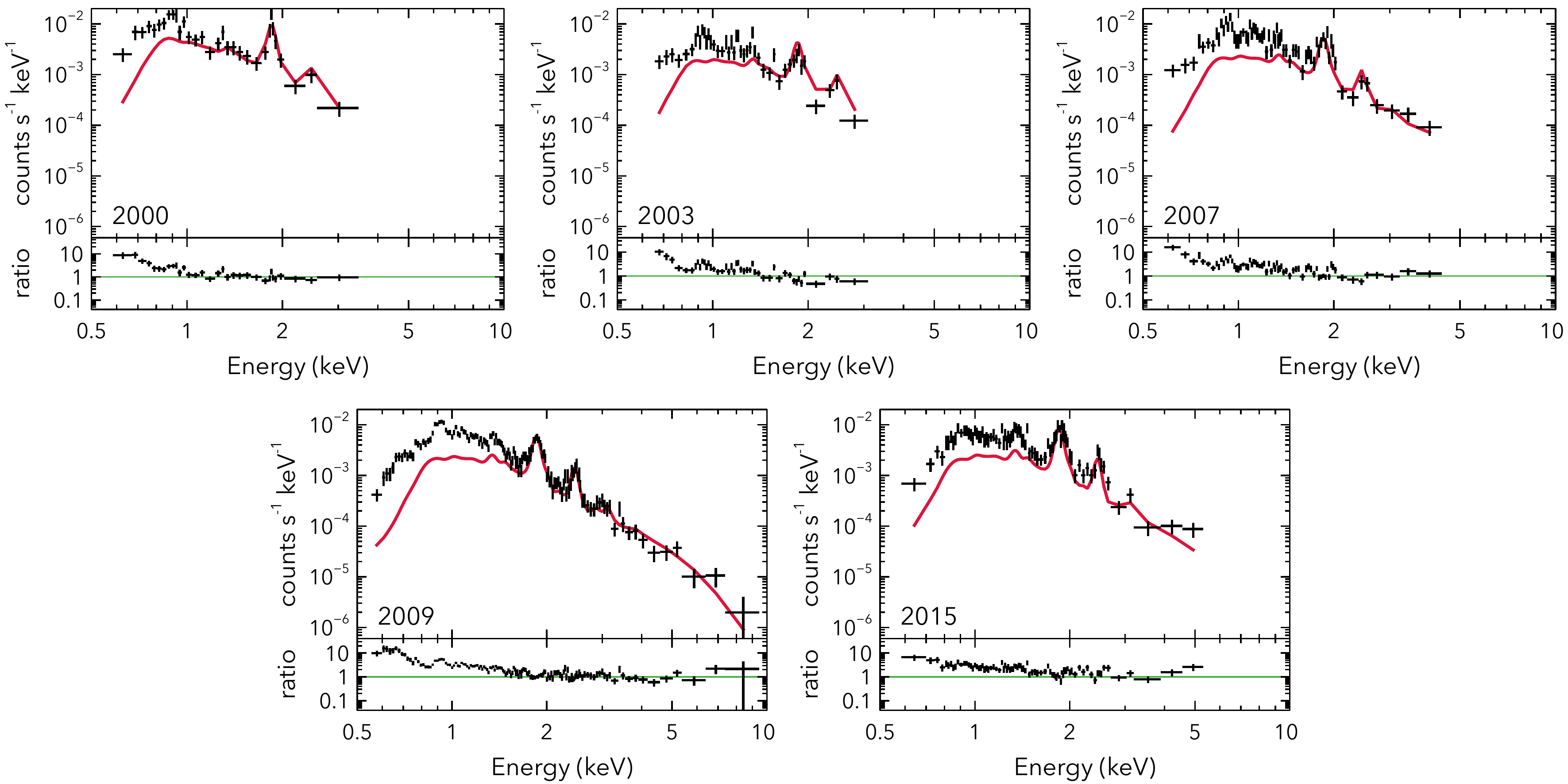}
	\caption{Comparison of the spectra of Knot1 (black) and the best-fit models of Ref1 for the effective area of Knot1 (red) in 2000, 20003, 2007, 2009, and 2015.
	The model is composed of a NEI component and a power-law component as explained in the text.
	Lower panels in each box show the ratio of the data of Knot1 to the model of Ref1.
	\label{fig:spec_ratio}}
	\vspace{0.7cm}
	\plotone{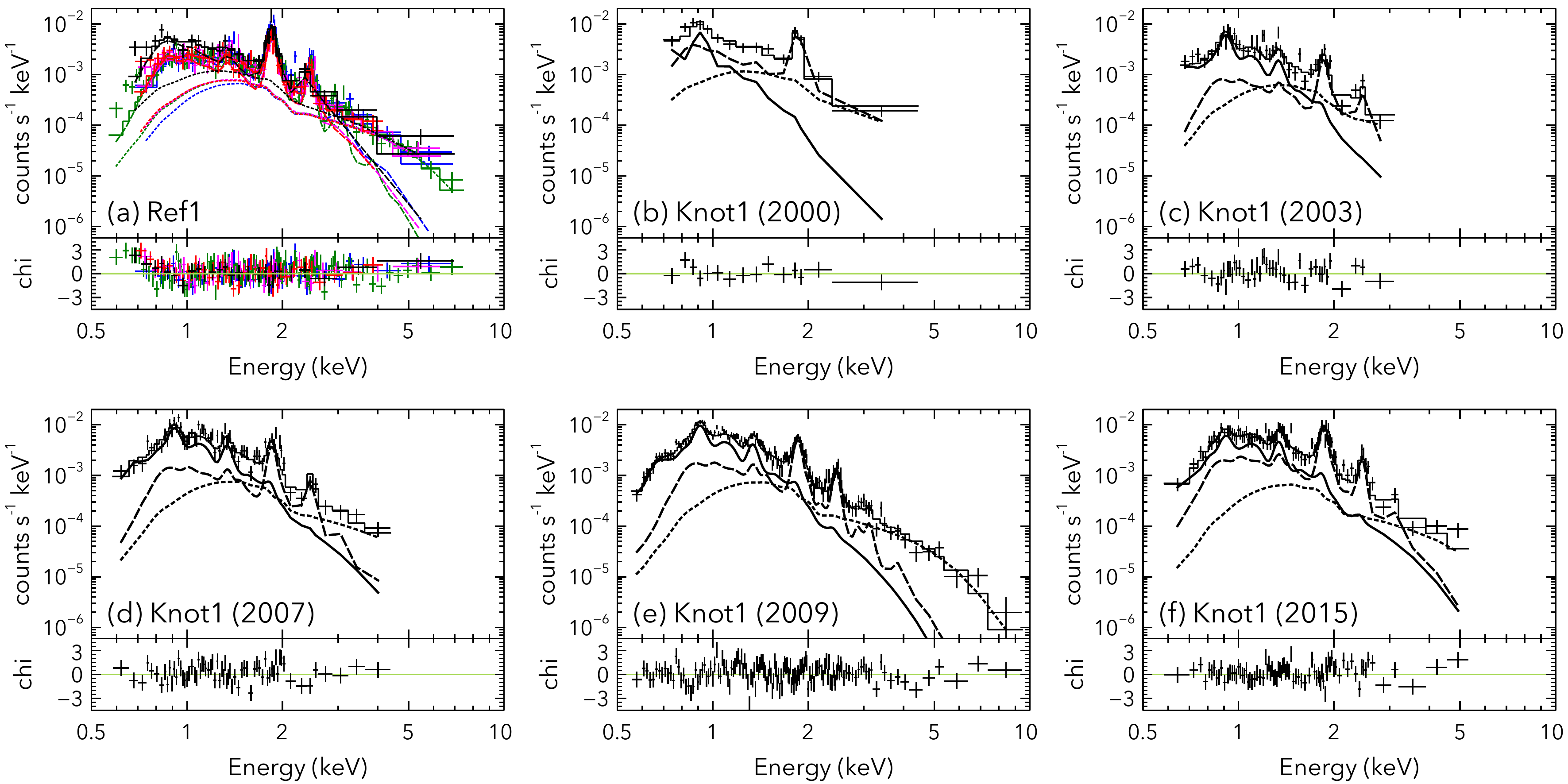}
	\caption{(a) Spectra of Ref1 taken in 2000 (black), 2003 (red), 2007 (magenta), 2009 (green), and 2015 (blue).
	The best-fit models are shown as the stepped lines.
	The dashed line and dotted line are the NEI component and non-thermal component, respectively.
	(b--f) Spectra of Knot1 taken in 2000, 2003, 2007, 2009, and 2015.
	The solid lines is the soft components.
	The other lines represent the same components as panel (a).
	\label{fig:spec}}
\end{figure*}

\subsection{Spectral Analysis}\label{subsec:spec}

In order to investigate the nature of Knot1 and quantitatively measure its time variability, we analyze spectra extracted from the region shown in Figure~\ref{fig:bandimg}.
The datasets obtained in each year are merged; we thus obtain five spectra from five different epochs (2000, 2003, 2007, 2009, and 2015).
A background spectrum is extracted from a blank region outside the remnant.
To estimate the contribution from the emission in the energy above the middle band in Knot1, we also extract spectra from a nearby reference region (noted as ``Ref1'' in Figure~\ref{fig:bandimg}), where the flux of the middle band component is almost the same as Knot1.

Comparing the Knot1 spectra with the best-fit model of Ref1, we reveal that the Knot1 emission in $\lesssim1.5$~keV band is significantly brighter than that of Ref1 while not in $\gtrsim1.5$~keV band (Figure~\ref{fig:spec_ratio}).
The thermal radiation in Ref1 is likely to have the same origin of the southeastward diffusing ejecta since the ejecta emission generally dominates the thermal radiation in the most inner region of the remnant \cite[e.g.,][]{cass07,mice15}.
This is supported by its spectrum which can be reproduced by a pure-metal NEI model. 
Thus, it is plausible to interpret that the excess emission of Knot1 in the soft band is due to the radiation from the knot structure.
The figure reveals that the energy band with the high Knot1/Ref1 ratio extends to higher energy year by year, supporting the flux increase in the soft band of Knot1.
To estimate the time variability in the soft band more quantitatively, we model the spectra of Knot1 with a soft component added to the Ref1 model.

We simultaneously fit the spectra of the Knot1 and Ref1 region taken in 2000, 2003, 2007, 2009, and 2015.
The spectrum of Ref1 is modeled with an absorbed non-equilibrium ionization (NEI) plus a power-law according to previous studies \citep[e.g.,][]{yama17}.
For the following analysis, we use version 12.10.1f of XSPEC \citep{arna96} with AtomDB version 3.0.9 \citep{fost17}.
The NEI components represent the ejecta of the remnant.
The electron temperatures ($kT_e$), ionization age ($n_e t$), and abundances of these components are assumed to be shared for each year.
Because Tycho is a remnant of Type Ia SN, H, He, and N are assumed to be absent in the ejecta.
O and Ne are fixed at solar value with respect to C, whose atomic number is the lowest in the element of the ejecta.
The abundance of the other elements is free.
The emission measure (EM) is defined as $\int n_e n_\mathrm{C} dV/4\pi d^2\mathrm{[C/H]_\sun}$, where $n_e$ and $n_\mathrm{C}$ are the number densities of electrons and carbon, and $V$ is the volume of the emitting plasma.
The photon index of the power law is fixed to 2.79, which is the value obtained from a nearby non-thermal-dominated region.
The normalization of the power laws in 2000, 2003, 2007, 2009, and 2015 are linked to each other.
We applied the T\"{u}bingen-Boulder model \citep{wilm00} for the interstellar absorption.
The result of the spectral fit and best-fit parameters of Ref1 region are shown in Figure~\ref{fig:spec}~(a) and Table~\ref{tab:parameters}, respectively.

\begin{deluxetable*}{llccccc}
	\tablecaption{Best-fit Parameters of Ref1 and Knot1 regions.\label{tab:parameters}}
	\tabletypesize{\footnotesize}
	\tablehead{
	\colhead{Components} & \colhead{Parameters (Units)} & \colhead{2000} & \colhead{2003} & \colhead{2007} & \colhead{2009} & \colhead{2015}
	}
	\startdata
	Absorption  & $N_\mathrm{H}$~($10^{22}$~cm$^{-2}$) & \multicolumn{5}{c}{$1.01_{-0.03}^{+0.04}$} \\
	\hline
		Ref1 region &&&\\
		NEI comp.		& EM\tablenotemark{$*$}~(10$^{9}$~cm$^{-5}$) & $2.4_{-1.1}^{+1.3}$ & $1.6_{-1.2}^{+1.4}$ & $2.3_{-1.2}^{+1.3}$ & $3.0_{-1.2}^{+1.4}$ & $5.2_{-2.0}^{+2.2}$ \\
			& $kT_e$~(keV) & \multicolumn{5}{c}{$0.70\pm0.03$} \\
			& $n_e t$~($10^{11}$~cm$^{-3}$~s) & \multicolumn{5}{c}{$5.0_{-1.2}^{+1.3}$} \\
			& [Mg/C]/[Mg/C]$_\sun$ & \multicolumn{5}{c}{$1.4_{-0.4}^{+0.8}$} \\
			& [Si/C]/[Si/C]$_\sun$ & \multicolumn{5}{c}{$11_{-3}^{+5}$} \\
			& [S/C]/[S/C]$_\sun$ & \multicolumn{5}{c}{$11_{-3}^{+6}$} \\
			& [Ar/C]/[Ar/C]$_\sun$($=$[Ca/C]/[Ca/C]$_\sun$) & \multicolumn{5}{c}{$9_{-4}^{+6}$} \\
			& [Fe/C]/[Fe/C]$_\sun$($=$[Ni/C]/[Ni/C]$_\sun$) & \multicolumn{5}{c}{$1.1_{-0.4}^{+0.8}$} \\
		Power law	& $\Gamma$ & \multicolumn{5}{c}{$2.79$ (fixed)} \\
			& Flux\tablenotemark{$\dag$}~($10^{-15}$~erg~cm$^{-2}$~s$^{-1}$) & \multicolumn{5}{c}{$1.7\pm0.1$} \\
		\hline
		Knot1 region &&&\\
		Soft comp.  & 
		EM\tablenotemark{$\ddag$}~(10$^{10}$~cm$^{-5}$) & $2.5_{-1.8}^{+2.2}$ & $1.6_{-0.6}^{+1.1}$ & $1.3_{-0.4}^{+0.6}$ & $1.8_{-0.3}^{+0.4}$ & $1.2_{-0.3}^{+0.4}$ \\
			& $kT_e$~(keV) & $0.30_{-0.07}^{+0.05}$ & $0.43\pm0.10$ & $0.57_{-0.10}^{+0.14}$ & $0.51\pm0.05$ & $0.69_{-0.12}^{+0.16}$ \\
			& $n_e t$~($10^{9}$~cm$^{-3}$~s) & 4.8 & $8.8_{-1.9}^{+12.0}$ & $7.5_{-1.4}^{+1.9}$ & $7.9_{-1.0}^{+1.3}$ & $7.6_{-1.5}^{+2.7}$ \\
			& abundance & \multicolumn{5}{c}{fixed to the solar value} \\
		Reference comp.\tablenotemark{$\P$} & EM\tablenotemark{$*$}~(10$^{9}$~cm$^{-5}$) & $2.4_{-0.9}^{+1.0}$ & $1.4_{-0.3}^{+0.7}$ & $2.3_{-0.9}^{+1.0}$ & $2.9_{-1.0}^{+1.1}$ & $5.2_{-2.1}^{+2.4}$ \\
		Power law	& $\Gamma$ & \multicolumn{5}{c}{2.79 (linked to Ref1)} \\
			& Flux\tablenotemark{$\dag$}~($10^{-15}$~erg~cm$^{-2}$~s$^{-1}$) & \multicolumn{5}{c}{1.7 (linked to Ref1)} \\
		\hline
		$\chi^{2}$~(d.o.f.) && \multicolumn{5}{c}{731~(629)} \\
		\hline
	\enddata
	\tablecomments{
		\tablenotetext{$*$}{EMs for the NEI components are defined as $\int n_e n_\mathrm{C}dV/(4\pi d^2\mathrm{[C/H]}_\sun)$.}
		\tablenotetext{$\dag$}{The energy flux in the energy band of 4--6~keV.}
		\tablenotetext{$\ddag$}{EMs for the soft components are defined as $\int n_e n_\mathrm{H}dV/4\pi d^2$.}
		\tablenotetext{$\P$}{The parameters of the reference components other than the EMs are linked to the NEI component for the Ref1 region.}}
\end{deluxetable*}

\begin{figure}
	\plotone{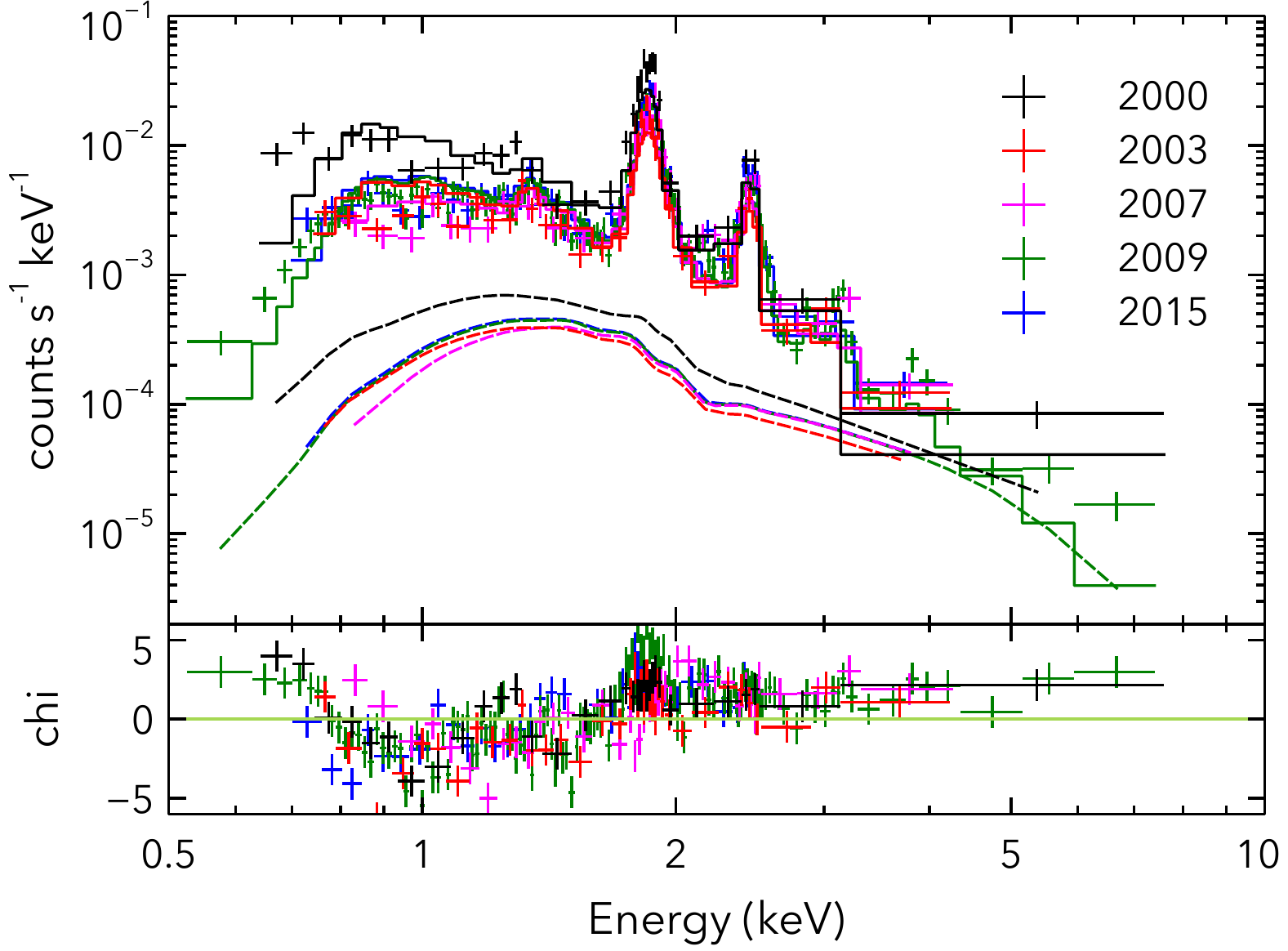}
	\caption{The spectra and best-fit model of Ref2 taken in 2000 (black), 2003 (red), 2007 (magenta), 2009 (green), and 2015 (blue).
	The lines show the model whose parameters except for EM is fixed to those of Ref1.
	The components represent the same ones as Figure~\ref{fig:spec}.
	}\label{fig:spec_ref}
\end{figure}

\begin{figure*}
	\plotone{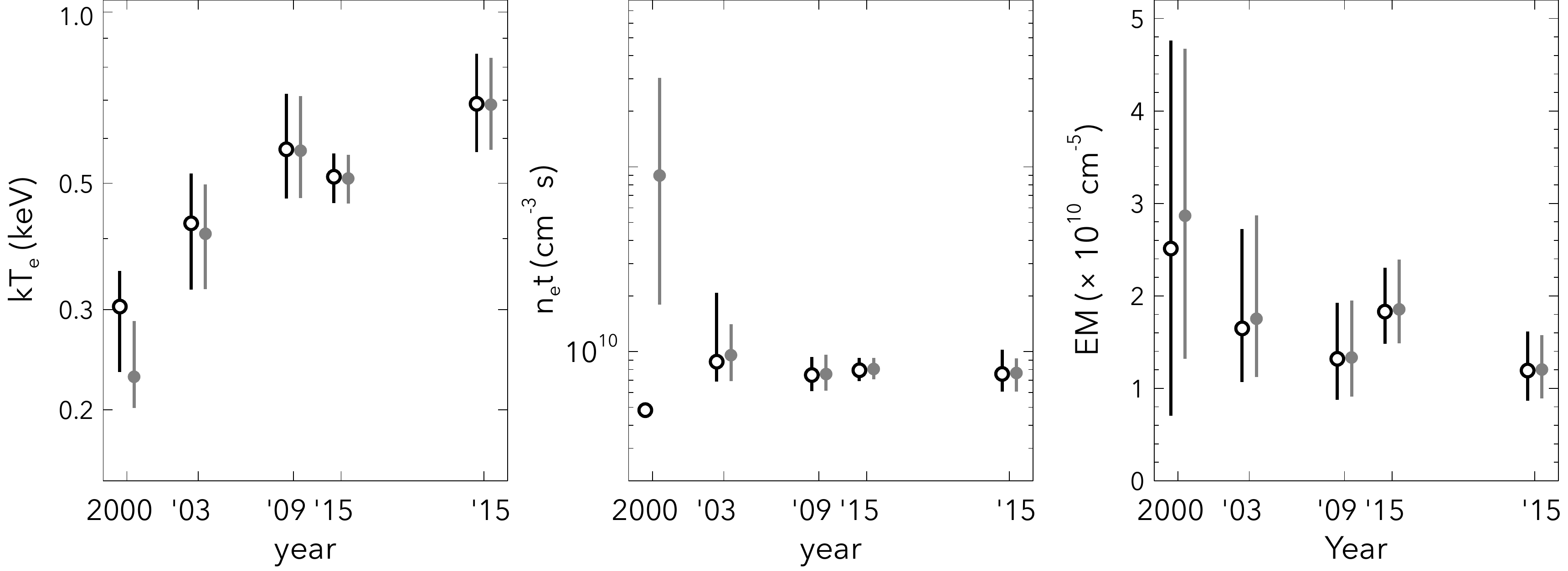}
	\caption{Time variations of $kT_e$ (left), $n_e t$ (middle), and EMs (right) of the best-fit soft component of Knot1. 
	The black and grey plots represent the best-fit parameters when $n_e t$ is fixed and free, respectively.
	\label{fig:para_vari}}
\end{figure*}

We fit the Knot1 spectra with a model consisting of the Ref1 component and an additional soft component.
The NEI model is used for the soft component.
The abundances of each element are fixed to the solar value.
Values of $n_e t$ in years other than 2000, $kT_e$, and EMs are set as free parameters.
Only $n_e t$ in 2000 cannot be determined because of a lack of statistics.
We thus fixed $n_e t$ in 2000 to that in 2003 minus $4\times10^9$~cm$^{-3}$~s ($=42$~cm$^{-3}\times3$~yr).
Note that fixing $n_e t$ does not change the other parameters beyond the 1$\sigma$ confidence level.
The EMs of the Ref1 component are free parameters, and the other parameters are linked to those for the Ref1 spectra.

\begin{figure*}[t]
	\plotone{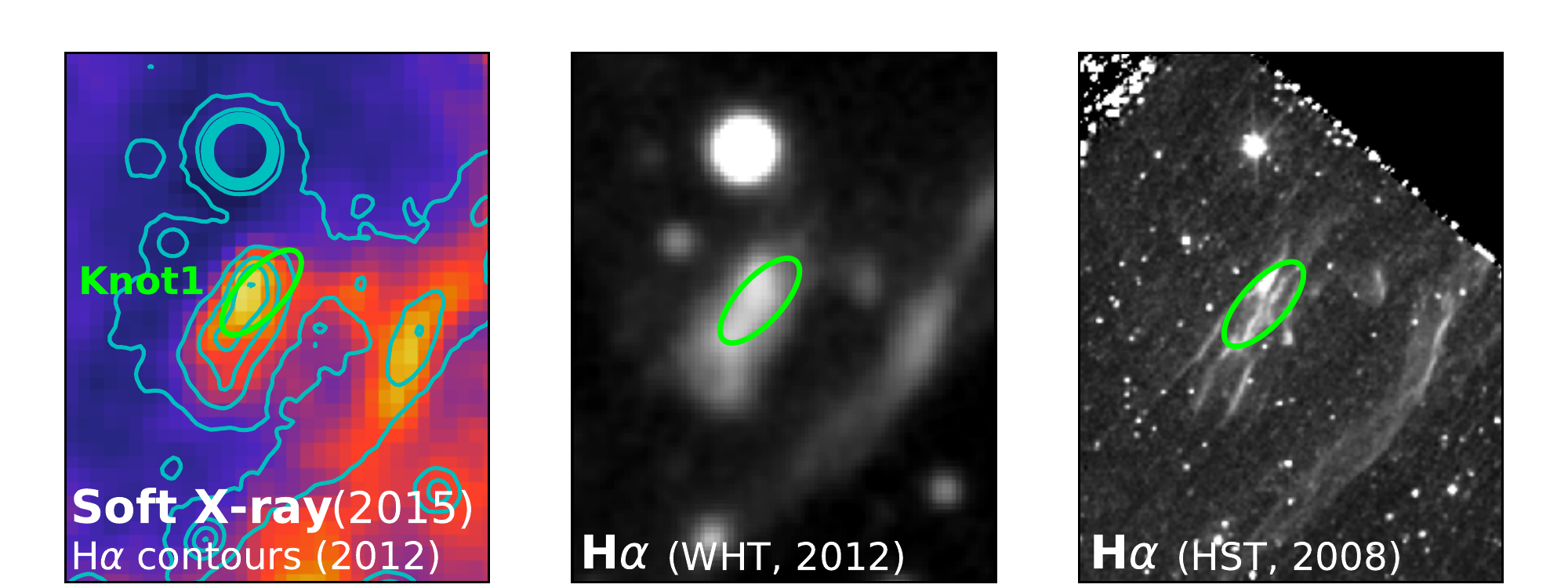}
	\caption{Left: Soft-band Chandra image taken in 2015 (same as the right-most panel of Figure \ref{fig:bandimg}) overlaid contours of an H$\alpha$ image taken in 2012 (see the middle panel).
	Middle: H$\alpha$ image obtained in 2012 with GH$\alpha$FaS on WHT \citep{knez17}.
	Right: H$\alpha$ image taken in 2008 with WFPC2 on HST \citep{lee10}.
	}\label{fig:correlation_H}
\end{figure*}

It may be possible that the uncertainty of $n_e t$ in 2000 is caused by contamination of X-rays from the southwest (SW).
We thus investigate a possibility of an extension of SW emission by checking a spectrum of an inner region of Knot1 toward the SW (the Ref2 region in Figure~\ref{fig:bandimg}).
Figure~\ref{fig:spec_ref} shows the Ref2 spectra and models whose parameters except for EM are fixed to those of Ref1.
As can be seen from the figure, the Ref2 spectra do not have the soft band excess like those of Knot1.
The result shows that the SW extension is negligible and that the soft thermal emission comes only from Knot1.

The spectra of Knot1 and the result of the spectral fit are presented in Figure~\ref{fig:spec}~(b--f).
The best-fit parameters are listed in Table~\ref{tab:parameters}.
We confirm that the time variability can be ascribed solely to the additional soft component.
Since the NEI model with the solar abundance can reproduce the Knot1 spectra in the soft band well, the soft component can be attributed to ISM heated up by the blast wave.
To further clarify the variability, we plot $kT_e$, $n_e t$, and EM as a function of time in Figure~\ref{fig:para_vari}.
We also show the parameters when $n_e t$ is a free parameter in the figure.
$kT_e$ increases significantly from $0.30_{-0.07}^{+0.05}$~keV to $0.69_{-0.12}^{+0.16}$~keV in 2000--2015.
As can be seen from the figure, $kT_e$ in both cases are almost equal.
We also confirm the $kT_e$ change when $n_e t$ is fixed to $8\times10^9$~cm$^{-3}$~s, which is between the best-fit value ($9\times10^{10}$~cm$^{-3}$~s) and the fixed value ($4\times10^9$~cm$^{-3}$~s).
In this case, $kT_e$ in 2000, 2003, 2007, 2009, and 2015 are $0.47_{-0.28}^{+0.20}$~keV, $0.42_{-0.04}^{+0.10}$~keV, $0.59_{-0.10}^{+0.13}$~keV, $0.53_{-0.06}^{+0.03}$~keV, and $0.70_{-0.13}^{+0.14}$~keV, respectively.

Based on the $kT_e$ increase, we also model the soft component with \verb|gnei|, an NEI model in which the ionization timescale averaged temperature is not necessary to be equal to the current temperature.
The value of $kT_e$ of \verb|gnei| are $0.26_{-0.05}^{+0.07}$~keV, $0.37_{-0.07}^{+0.09}$~keV, $0.57_{-0.10}^{+0.13}$,~keV $0.52_{-0.07}^{+0.05}$~keV, and $0.70_{-0.13}^{+0.17}$~keV in 2000, 2003, 2007, 2009, and 2015, respectively; they are almost the same as those of the \verb|nei| model.
We find no notable changes in $n_e t$ and EM over time.
We can interpret that the observed flux change is due to an increase of electron energy caused by the shock heating.

%%%%%%%%%%%%%%%%%%%%%%%%%%%%%%%%%%%%%%%%%%%%%%%%%%%%%%%%%%%%%%%%%%%%%%%%%%%%%%%%
\section{Discussion}\label{sec:discussion}
\subsection{Origin of Knot1}\label{sec:ism}

As described in Section~\ref{sec:ana}, the significant increase of the soft-band X-ray flux is seen in Knot1 in Tycho.
%The result together with the year-scale increase of the electron temperature implies that a compact dense clump was recently heated by the blast wave.
Together with the year-scale increase of the electron temperature, the result implies that a compact dense clump was recently heated by the blast wave.
The model with the solar abundance reproduces the spectra, suggesting that the shock-heated gas is of ISM origin (Table~\ref{tab:parameters}).
We do not, however, rule out the possibility of CSM origin since the southwestern shell is known to be interacting with a cavity wall \citep{tana21}.
Note that Knot1 is the first example of ISM/CSM X-ray emission in the ejecta-dominated SNR, Tycho.
Future observations with improved statistics will enable to measure the abundance of each element, resulting in determination of its true origin.
It may also hint at the progenitor system of Tycho's Type Ia SN.

In the northeastern region, previous H$\alpha$ observations \citep{kirs87} revealed Balmer-dominated filaments, which are interpreted as radiation from a forward-shocked neutral gas and shock precursors \citep[e.g.,][]{ghav00, lee07}.
Figure~\ref{fig:correlation_H} (the left and middle panels) shows a comparison between the soft-band X-ray image taken in 2015 and the H$\alpha$ image taken in 2012 \citep{knez17}.
We find that a bright H$\alpha$ structure spatially coincides with Knot1 in X-rays.
This finding supports the ISM/CSM origin of Knot1.
%While the clear spatial correlation indicates that these emissions may have the same origin, X-ray emission is generally expected behind a shock front.
%We, therefore, speculate that the H$\alpha$ structure consists of multiple filaments, which is confirmed by another H$\alpha$ observation with better angular resolution \citep{lee10} as shown in the right panel of Figure~\ref{fig:correlation_H}.

We point out that the bright and complicated shell structure is seen only around Knot1 in the entire H$\alpha$ image of the northeastern part of Tycho taken with the Hubble Space Telescope \citep[the right panel of Figure~\ref{fig:correlation_H} and cf.][]{lee10}.
Similar localized multiple filaments are present in other SNRs; the ``XA'' region and the southwestern limb of the Cygnus Loop \citep{hest86,grah95} and an ejecta knot of Cas~A \citep{patn07, patn14}.
In the case of Cas~A, time variability of thermal X-rays was detected in a physical scale of 0.02--0.03~pc, which roughly agrees with the estimated size of Knot1: $\simeq0.04$~pc.
These structures are considered as dense clumps engulfed by the blast waves.
We thus infer that Knot1 originate from a small-scale clumpy ISM/CSM heated by the forward shock.

Here we estimate the density of Knot1 using the best-fit parameters as follows.
Assuming that the emitting region of Knot1 has an oblate-spheroidal shape with long and short radii of 0.05~pc and 0.02~pc, respectively, we obtain its volume of $V\simeq6\times10^{51}$~cm$^{3}$.
From the best-fit parameter of the soft component in 2015, the emission measure is $n_e n_\mathrm{H}V/4\pi d^2=(1.2_{-0.3}^{+0.4})\times10^{10}$~cm$^{-5}$, from which we derive a proton density of $n_\mathrm{H}=35_{-4}^{+6}$~cm$^{-3}$.
Since the post-shock density of Tycho is estimated to be $n_\mathrm{H}=0.1$--2~cm$^{-3}$ from the flux ratio of the 70~$\mu$m to 24~$\mu$m infrared emission \citep{will13}, the small clump in Knot1 has roughly 10--100 times higher density than the surroundings.
%%%%%%%%%%%%%%%%%%%%%%%%%%%%%%%%%%%%%%%%%%%%%%%%%%%%%%%%%%%%%%%%%%%%%%%%%%%%%%%%

\subsection{Time Variability of Knot1}\label{sec:vari}
\subsubsection{Cloud Crushing Time}\label{sec:cct}
\begin{figure}
	\plotone{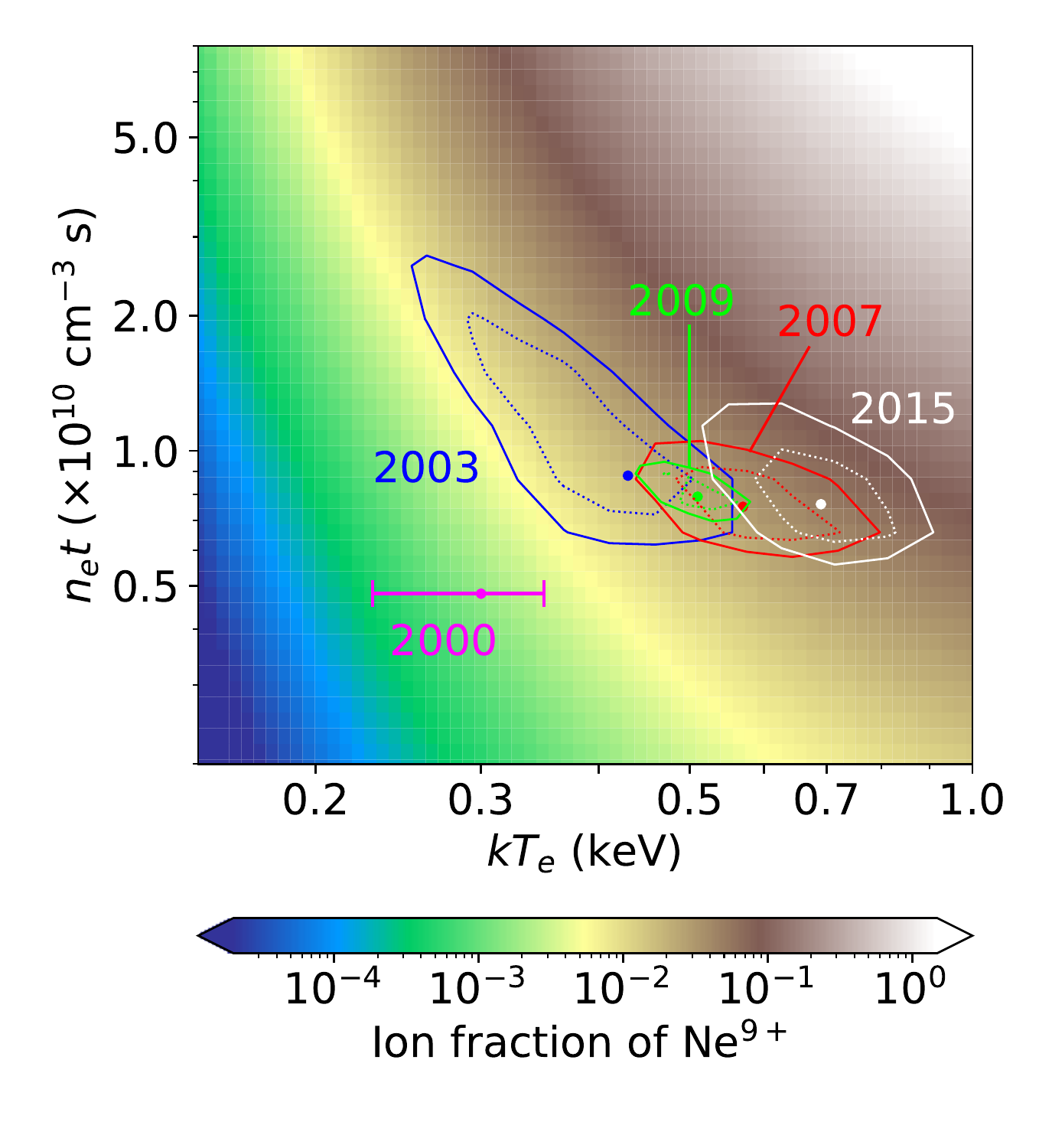}
	\plotone{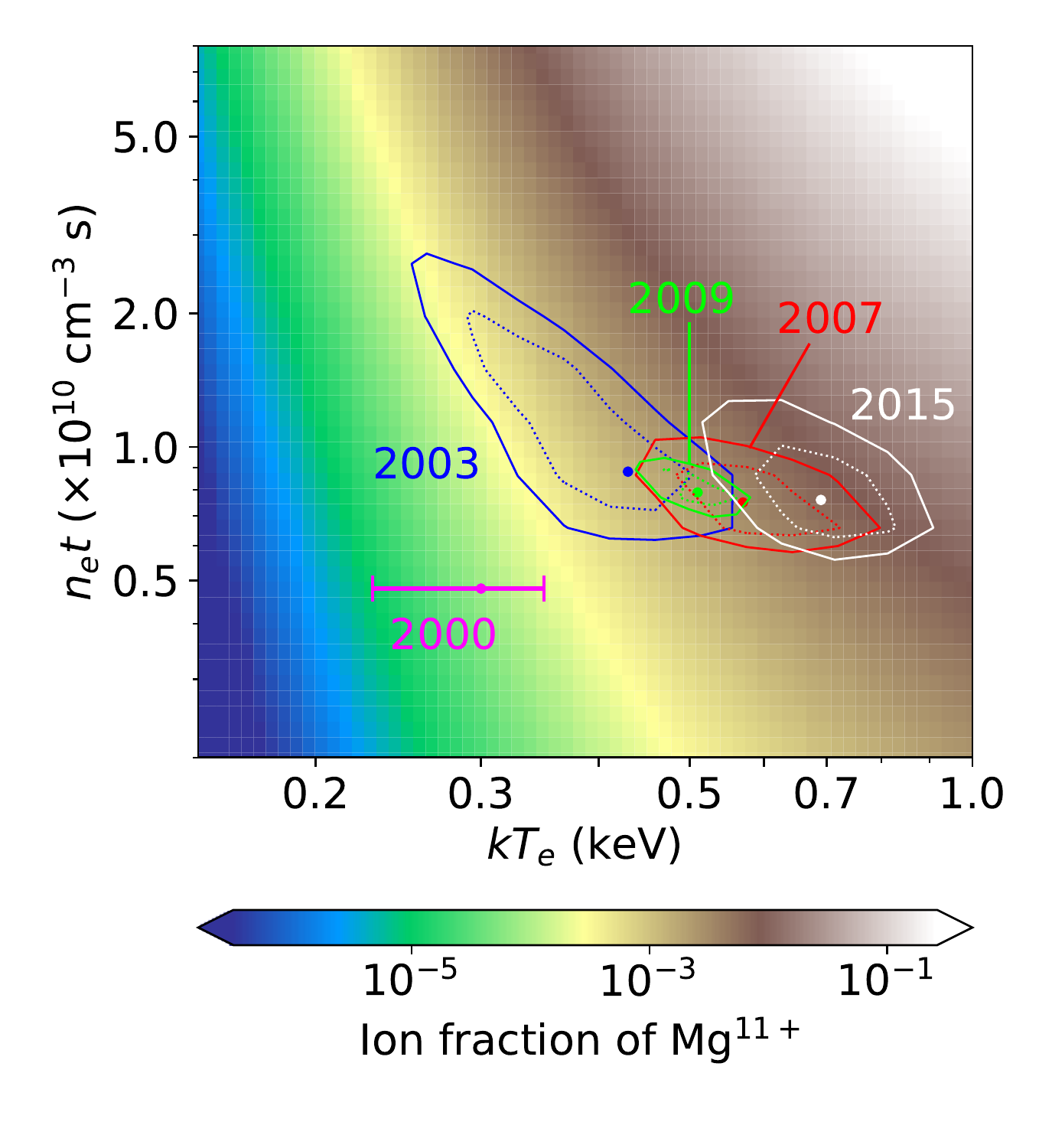}
	\caption{
	Ion fractions of Ne$^{9+}$ (top) and Mg$^{11+}$ (bottom), i.e., H-like ions , as a function of $kT_e$ and $n_e t$.
	The dotted and solid line contours show $\Delta\chi^2=1.0$ and 2.3 confidence levels, respectively, in 2003 (blue), 2007 (red), 2009 (green), and 2015 (white).
	The points represent the best-fit parameters in each year.
	For 2000 data, we show only the uncertainty of $kT_e$ since $n_e t$ was fixed in our analysis (see text).
	\label{fig:Hfraction}}
\end{figure}

Since the parameter $n_{e}t$ in XSPEC represents the ionization timescale assuming constant $kT_e$, it is not reasonable, when $kT_e$ is significantly increasing, to consider $n_e t$ as a product of density and actual time passed.
In order to discuss the change in ionization state of Knot1, we calculated ion fractions of the soft component in each year.
As shown in Figure~\ref{fig:Hfraction}, H-like Ne and Mg are both increasing, supporting that the ionization has progressed from 2003 to 2015.
We thus consider that Knot1 is heated and ionized year to year by an SNR shock recently propagating into a small cloud.

To estimate the timescale for shock heating, we assume ram pressure equilibrium ($\rho_\mathrm{i}u_\mathrm{i}^2\simeq\rho_\mathrm{c}u_\mathrm{c}^2$), where $\rho$ and $u$ are the density and velocity, respectively, in ISM (subscript i) and inside the clump (subscript c).
The velocity of the shock decelerated inside the clump is described as 
\begin{equation}
u_\mathrm{c}=\frac{u_\mathrm{i}}{\chi^{1/2}}.
\end{equation}
Here, $\chi$ ($\equiv\rho_\mathrm{c}/\rho_\mathrm{i}$) is the density contrast between the clump and ISM.
Assuming $\chi=n_\mathrm{c}/n_\mathrm{i}\simeq10$ following the discussion in Section~\ref{sec:ism} and that the forward-shock velocity ($u_\mathrm{i}$) is typical of Tycho \citep[4000--8000~\kms;][]{tana21}, we obtain $u_\mathrm{c}=1500$--2500~\kms.

Following the discussion by \citet{patn14}, we define a cloud crushing time:
\begin{equation}
	t_\mathrm{cc}\equiv\frac{\chi^{1/2}a_0}{u_\mathrm{i}}=\frac{a_0}{u_\mathrm{c}},
\end{equation}
where $a_0$ is the radius of the clump \citep{klei94}.
Since the radius of the X-ray emitting region of Knot1 is $a_0\simeq0.02$~pc, the cloud crushing time is $t_\mathrm{cc}=18\times(u_\mathrm{i}/2000~\mathrm{km~s^{-1}})^{-1}~\mathrm{yr}$.
We point out that the result is roughly consistent with the year-scale change of the X-ray flux in Knot1.
%%%%%%%%%%%%%%%%%%%%%%%%%%%%%%%%%%%%%%%%%%%%%%%%%%%%%%%%%%%%%%%%%%%%%%%%%%%%%%%%

\subsubsection{Heating Timescale}
\begin{figure*}
    \epsscale{1.1}
	\plotone{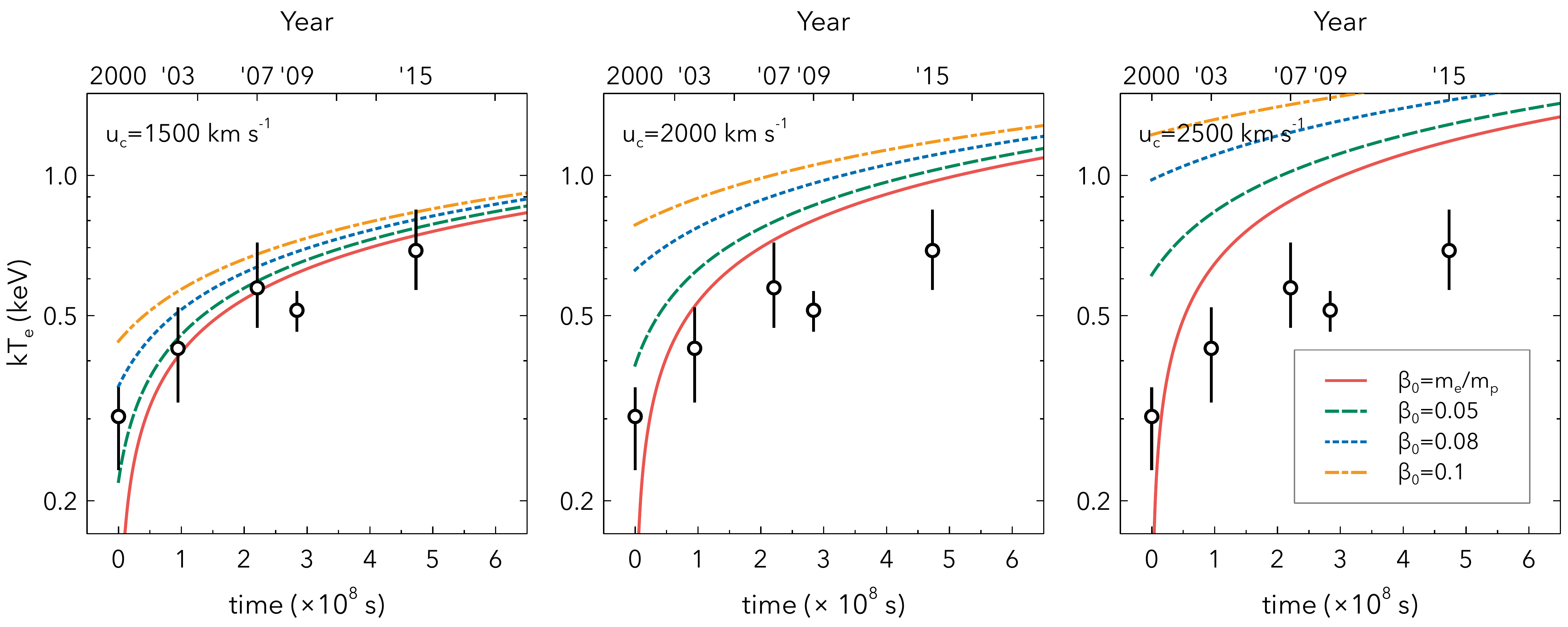}
	\caption{
	Best-fit results of $kT_e$ for Knot1 (same as the left panel of Figure \ref{fig:para_vari}) and calculated  time variations  with assumptions of 
	$u_\mathrm{c}=1500$~\kms \ (left), 2000~\kms \ (middle), 2500~\kms \ (right).
	The solid, dashed, dotted, and dash-dotted lines represent cases of $\beta_0=m_e/m_p\simeq1/2000$, 0.05, 0.08, and 0.1, respectively.
	\label{fig:kte_uc}}
	\epsscale{0.8}
	\plotone{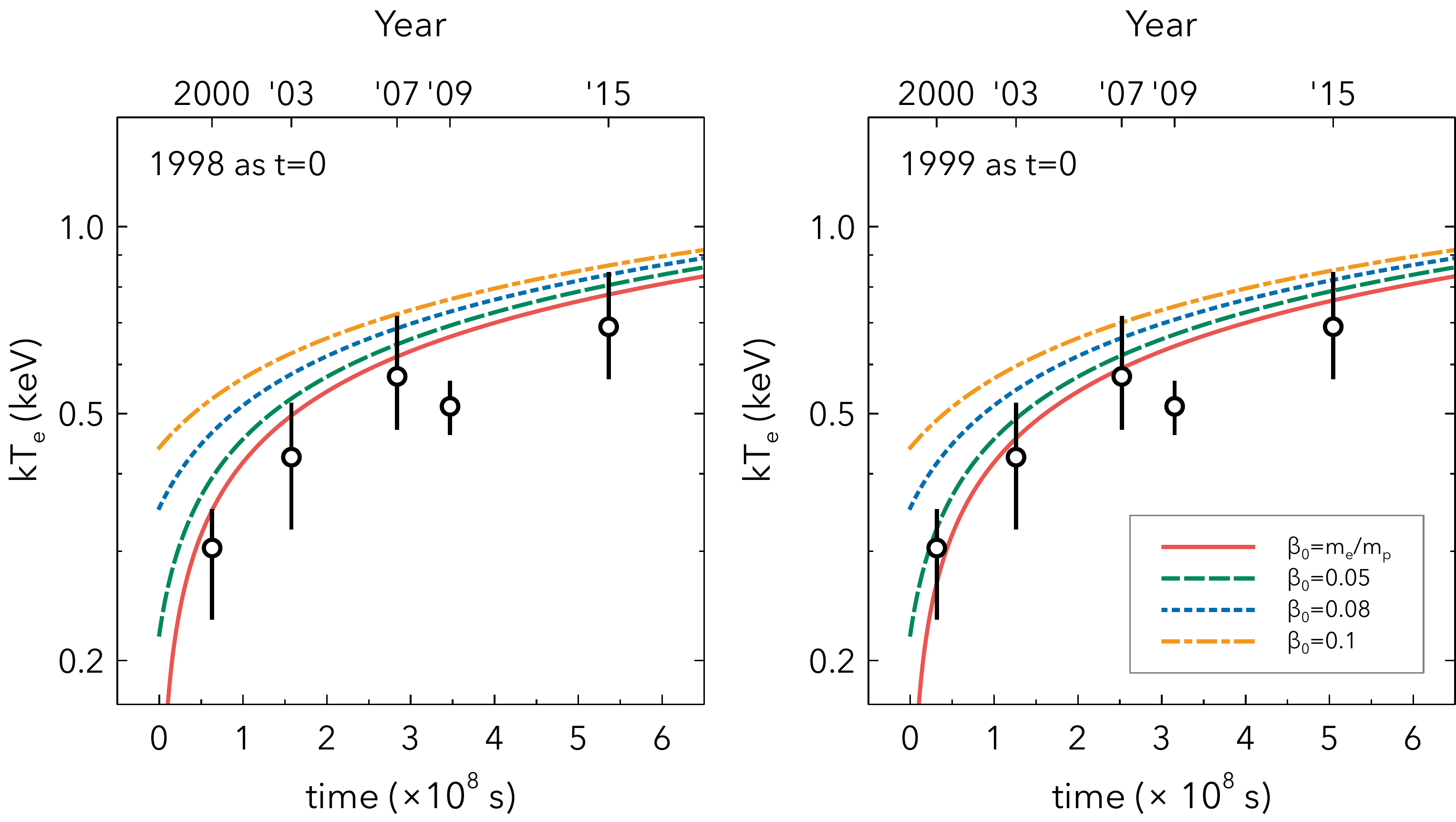}
	\caption{
	Same as Figure~\ref{fig:kte_uc}, but assuming the years 1998 (left) and 1999 (right) as $t=0$ in the case of $u_\mathrm{c}=1500$~\kms.
	\label{fig:kte_t0}}
\end{figure*}

To explain the observed increase of $kT_e$ in Knot1, we first assume a thermal equilibration via ion-electron Coulomb collisions without collisionless heating at the shock transition region.
The immediate downstream temperature for a shock velocity $u_\mathrm{c}$ is written as
\begin{equation}\label{eq:RH}
	kT_i=\frac{3}{16}m_i u_\mathrm{c}^{2},
\end{equation}
where $m_i$ is the mass of particle species $i$.
Since the electron temperature ($T_e$) is lower than the ion temperature ($T_i$) in the downstream plasma, a simple increase of $T_e$ is expected, and its time evolution is described as
\begin{equation}
	\frac{dT_e}{dt}=\frac{T_i-T_e}{t_\mathrm{eq}},
	\label{eq:kte}
\end{equation}
where the equilibration timescale $t_\mathrm{eq}$ is given by the following expression \citep{spit62,masa84}:
%
%\begin{eqnarray}
%	t_\mathrm{eq}&=&3.1\times10^8 \langle A_i \rangle\langle z \rangle^{-2} \nonumber \\
%		&&\quad \times\left( \frac{kT_e}{\mathrm{eV}} \right)^{3/2}\left(\frac{n_e}{\mathrm{cm}^{-3}}\right)^{-1}(\ln{\Lambda})^{-1}~\mathrm{s}, \\
%	\ln{\Lambda}&=&24.8+\ln{\left[ \left(\frac{kT_e}{\mathrm{eV}}\right)\left(\frac{n_e}{\mathrm{cm}^{-3}}\right)^{-1/2} \right]}.
%\end{eqnarray}
%
\begin{eqnarray}
	t_\mathrm{eq}&=&\frac{3m_e m_i k^{3/2}}{8(2\pi)^{1/2}nZ_{i}^{2}e^{4}\ln{\Lambda}}\left(\frac{T_e}{m_e}+\frac{T_i}{m_i}\right)^{3/2} \\
	\ln{\Lambda}&=&24.8+\ln{\left[ \left(\frac{kT_e}{\mathrm{eV}}\right)\left(\frac{n_e}{\mathrm{cm}^{-3}}\right)^{-1/2} \right]}.
\end{eqnarray}
Here, $Z_i$ and $e$ are the charge number and the elementary charge, respectively.
We take the  electron number density $n_e=42$~cm$^{-3}$ under the assumption of  $n_\mathrm{H}=35$~cm$^{-3}$ (Section~\ref{sec:ism}) and $n_e=1.2n_\mathrm{H}$.
Assuming that no contributions from ions heavier than hydrogen for simplicity, $kT_e$ evolves as shown in Figure~\ref{fig:kte_uc}.
If only Coulomb collisions are considered, the electron-to-proton temperature ratio ($\beta_{0}\equiv T_e/T_p$) at $t=0$  should be equal to the mass ratio of the particles, i.e., $\beta_0=m_e/m_p\simeq5\times10^{-4}$.
From Figure~\ref{fig:kte_uc}, we find that the model for $u_\mathrm{c}=1500$~\kms \  can explain the result.

\begin{figure*}
    \epsscale{1.1}
	\plotone{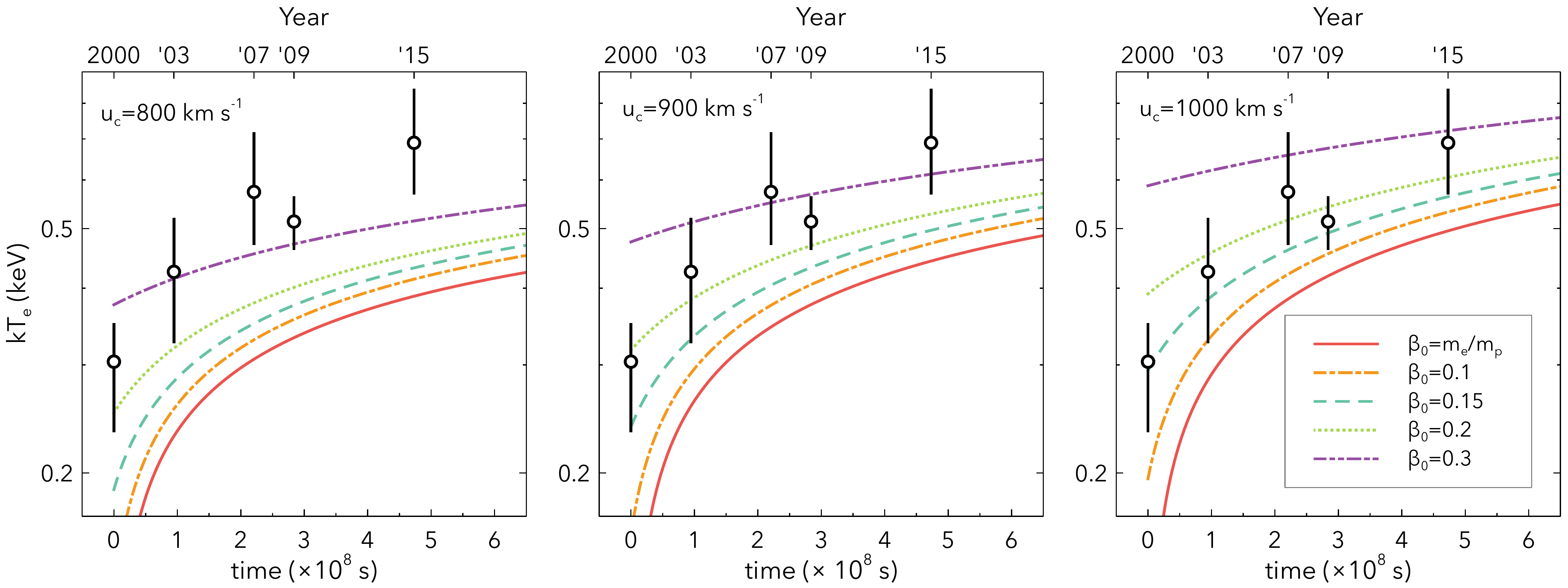}
	\caption{
	Best-fit results of $kT_e$ for Knot1 (same as the left panel of Figure \ref{fig:para_vari}) and calculated  time variations  with assumptions of 
	$u_\mathrm{c}=800$~\kms \ (left), 900~\kms \ (middle), 1000~\kms \ (right).
	The solid, dash-dotted, dashed, dotted, and dash-dot-dotted lines represent cases of $\beta_0=m_e/m_p\simeq1/2000$, 0.1, 0.15, 0.2 and 0.3, respectively.}
	\label{fig:kte_uc_lowv}
	\epsscale{0.8}
	\plotone{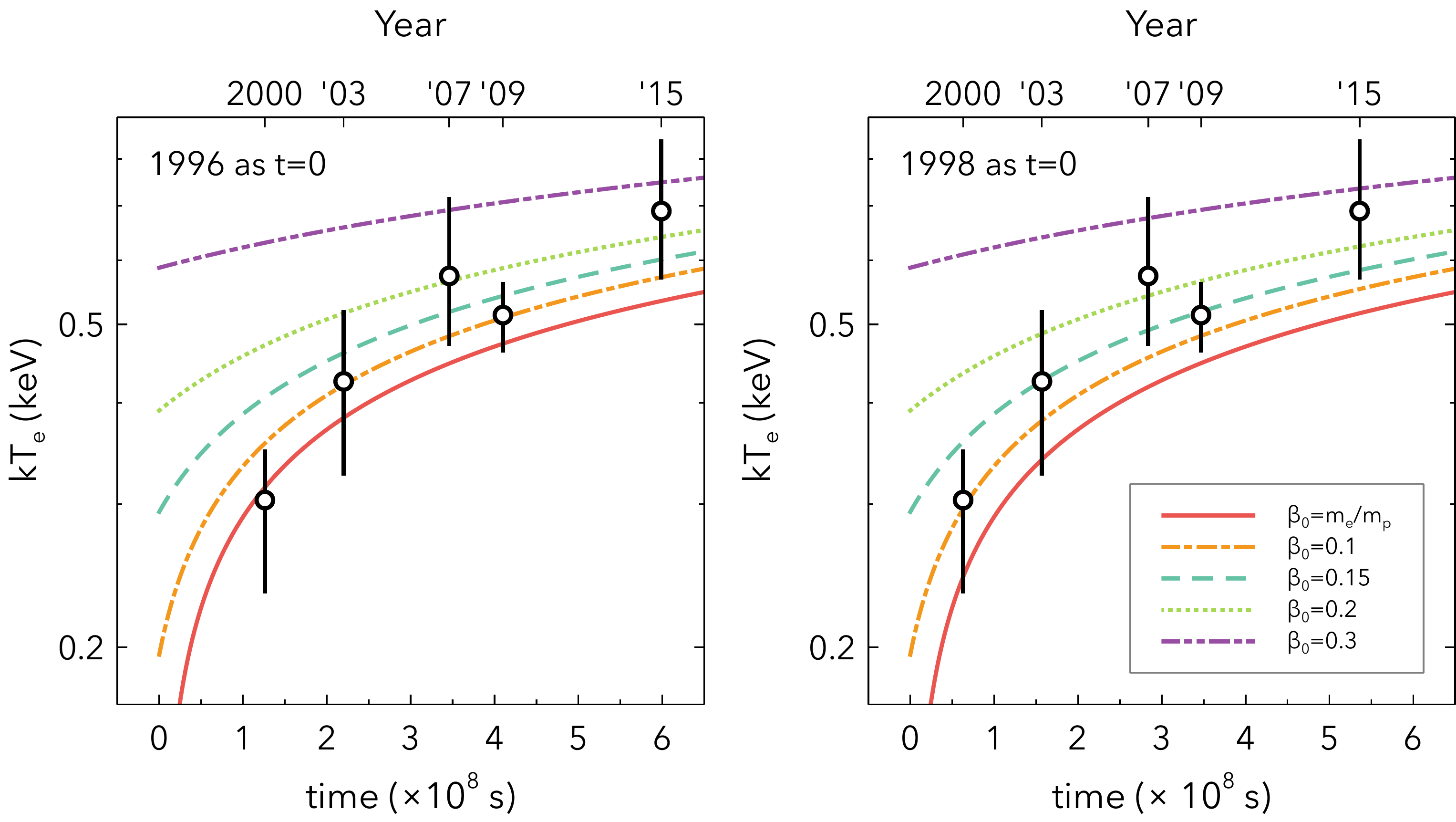}
	\caption{
	Same as Figure~\ref{fig:kte_uc_lowv}, but assuming the years 1996 (left) and 1998 (right) as $t=0$ in the case of $u_\mathrm{c}=1000$~\kms.
	\label{fig:kte_t0_lowv}}
\end{figure*}

When the collisionless process is effective at the shock transition \citep[e.g.,][]{carg88, lami00, ghav07}, the ratio $\beta_0$ should be larger than $m_e/m_p$ ($\simeq5\times10^{-4}$).
We try $\beta_0=0.05$, 0.08, and 0.1 as plotted in Figure~\ref{fig:kte_uc}.
The model with $m_\mathrm{e}/m_\mathrm{p}\leq\beta_{0}\leq0.05$ agrees well with the data.
%, implying a low equilibration, that is, little or no collisionless heating at the shock transition in Knot1.
Since we do not know when the forward shock indeed hit Knot1, we compare several calculations with different assumptions about $t=0$ in the case of $u_0=1500$~\kms \ in Figure~\ref{fig:kte_t0}.
Even in the case of the year 1998 as $t=0$, $m_\mathrm{e}/m_\mathrm{p}\leq\beta_{0}\leq0.05$ is still possible, but the year 1999 as $t=0$ is more plausible.

While the shock velocity $u_\mathrm{c}$ plausibly ranges from 1500~\kms \ to 2000~\kms \ as estimated in Section~\ref{sec:cct}, the  $kT_e$ trend may indicate a lower $u_\mathrm{c}$ than 1500~\kms.
%Considering the uncertainties of $\chi$ and $u_\mathrm{i}$, $u_\mathrm{c}$ could be lower than 1500~\kms.
Figure~\ref{fig:kte_uc_lowv} and \ref{fig:kte_t0_lowv} shows calculated time variations of $kTe$ in the case of $u_\mathrm{c}\leq1000$~\kms.
The observed $kT_e$ can be roughly explained for $m_\mathrm{e}/m_\mathrm{p}\leq\beta_{0}\leq0.15$.
If this slower $u_\mathrm{c}$ is the case, the density contrast  $\chi$ might be larger or the forward-shock velocity $u_\mathrm{i}$ might be slower than expected, which should be constrained by future observations.

Based on flux ratios of the broad-to-narrow components of the H$\alpha$ line, $\beta_0$ is estimated for some SNRs with different shock velocities \citep[e.g.,][]{adel08}.
In Tycho, \citet{ghav01} and \citet{adel08} obtained $\beta_0<0.1$ and $\beta_0=0.046^{+0.007}_{-0.006}$, respectively, in a well-known region ``knot~g'', located $\sim2\arcmin$ southeast of Knot1.
Other SNRs such as SN~1006 and Kepler's SNR, which have strong shocks with $v_\mathrm{sh}>1000$~\kms, have $\beta_0\sim0.05$ \citep{fese89,ghav02}.
On the other hand,  $\beta_0$ is greater than 0.1 in SNRs with slow shocks ($v_\mathrm{sh}\leq1000$~\kms): e.g., Cygnus Loop, RCW~86 \citep{ghav01}, and SNR~0548$-$70.4 \citep{smit91}.
Knot1 has $m_\mathrm{e}/m_\mathrm{p}\leq\beta_{0}\leq0.05$ in a shock velocity of about 1500~\kms, or $m_\mathrm{e}/m_\mathrm{p}\leq\beta_{0}\leq0.15$ in a shock velocity of about 1000~\kms, which is consistent with these previous studies.
%It indicates $\beta_0\sim0.05$ for high shock velocities regardless of surrounding environments, including ambient gas densities.
It indicates that Knot1 has collisionless electron heating with efficiency comparable to the result of the H$\alpha$ observation.
%%%%%%%%%%%%%%%%%%%%%%%%%%%%%%%%%%%%%%%%%%%%%%%%%%%%%%%%%%%%%%%%%%%%%%%%%%%%%%%%

\section{Conclusion\label{sec:conclusion}}

We searched for a short-timescale variability of thermal X-ray radiation in Tycho, using the Chandra X-ray Observatory data in 2000, 2003, 2007, 2009, and 2015.
We discovered a significant brightening of a compact emission in the northwestern limb (Knot1).
Our spectral analysis indicated that the time variability of Knot1 was due to a change of the electron temperature $kT_e$ of forward-shocked gas.
Knot1 was the first detection of shock-heated ISM/CSM in this remnant.
The best-fit result indicated a gradual increase of $kT_e$ from $0.30_{-0.07}^{+0.05}$~keV to $0.69_{-0.12}^{+0.16}$~keV of Knot1 during 2000--2015.
From these results, together with localized multiple H$\alpha$ filaments in Knot1, we considered that a small ($\simeq0.04$~pc in diameter)  dense ($n_\mathrm{H}\sim30$~cm$^{-3}$) clump  was recently encountered by the forward shock.
By calculating equilibration timescales of $kT_e$, $\beta_0$ ($\equiv T_e/T_p$) was required to be $m_\mathrm{e}/m_\mathrm{p}\leq\beta_{0}\leq0.05$ when shock velocity is 1500~km~s$^{-1}$ and $m_\mathrm{e}/m_\mathrm{p}\leq\beta_{0}\leq0.15$ when shock velocity is 1000~km~s$^{-1}$ to reproduce the observed change in the electron temperature.
Our result shows the collisionless heating in Knot1, which have comparable efficiency to the previous H$\alpha$ observations of knot~g in Tycho and the other SNRs with high shock velocities.
%%%%%%%%%%%%%%%%%%%%%%%%%%%%%%%%%%%%%%%%%%%%%%%%%%%%%%%%%%%%%%%%%%%%%%%%%%%%%%%%

%\vspace{7mm}
\acknowledgments
We thank Dr. Sladjana Kne{\v{z}}evi{\'c} for providing the WHT H$\alpha$ image.
We acknowledge the anonymous referee for useful comments and suggestions that improved the quality of this paper.
This research has made use of data obtained from the Chandra Data Archive and software provided by the Chandra X-ray Center (CXC) in the application package CIAO.
This work is supported by JSPS KAKENHI Scientific Research Grant Numbers JP21J20027 (M.M.), JP19K03915, JP22H01265 (H.U.), JP19H01936 (T.T.), JP22K18721, JP22H04572 (T.G.T.) JP21H04493 (T.G.T. and T.T.).
\facilities{CXO, ING:Herschel (GH$\alpha$FaS)}
\software{CIAO \citep{frus06},
	XSPEC \citep{arna96},
	SAOImageDS9 \citep{joye03}
}

\bibliography{matsuda_tycho_thermal}{}
\bibliographystyle{aasjournal}

\end{document}